\documentclass[twocolumn]{emulateapj}
\journalinfo{{\sc Submitted to The Astrophysical Journal}}
\usepackage{natbib}
\usepackage{amssymb}
\usepackage{graphicx}
\newcommand{\igr}{EXO 1722-363}

\begin{document}
\title{The Orbit of the Eclipsing X-ray Pulsar EXO 1722-363}
\author{
Thomas W. J. Thompson\altaffilmark{1},
John A. Tomsick\altaffilmark{1},
J. J. M. in 't Zand\altaffilmark{2},
Richard E. Rothschild\altaffilmark{1},
Roland Walter\altaffilmark{3}}
\altaffiltext{1}{Center for Astrophysics and Space Sciences, University of California, San Diego, La Jolla, CA 92093; email: tthompson@physics.ucsd.edu }
\altaffiltext{2}{Space Research Organization of the Netherlands, Sorbonnelaan 2, 3584 CA Utrecht, Netherlands}
\altaffiltext{3}{{\em INTEGRAL} Science Data Centre, Chemin d'Ecogia 16, 1290 Versoix, Switzerland}

\begin{abstract}
With recent and archival {\em Rossi X-Ray Timing Explorer} ({\em RXTE}) X-ray measurements of the heavily obscured X-ray pulsar EXO 1722-363 (IGR J17252-3616), we carried out a pulse timing analysis to determine the orbital solution for the first time. Using a single pulse period derivative, we connected datasets separated by over 2.4 years without pulse number ambiguity ($> 180,000$ pulses), implying continuous spin-up of the pulsar from 2003 August (and possibly earlier) until 2006 February (and possibly later). The orbital solution also shows that a torque reversal occurred sometime between 1998 November and 2003 August. The binary system is characterized by $a_{x} \sin{i} = 101 \pm 3$ lt-s and $P_{\rm orb} = 9.7403 \pm 0.0004$ days (90\% confidence), with the precision of the orbital period being obtained by connecting datasets separated by more than 7 years (272 orbital cycles). The orbit is consistent with circular, and $e < 0.19$ at the 90\% confidence level. The mass function is 11.7 $\pm$ 1.2 $M_{\sun}$ and confirms that this source is a High Mass X-ray Binary (HMXB) system. The orbital period, along with the previously known $\sim$414 s pulse period, places this system in the part of the Corbet diagram populated by supergiant wind accretors. Using previous eclipse time measurements by Corbet et al. and our orbital solution, combined with the assumption that the primary underfills its Roche lobe, we find $i > 61\degr$ at the 99\% confidence level, the radius of the primary is between 21 $R_{\sun}$ and 37 $R_{\sun}$, and its mass is less than about 22 $M_{\sun}$. The acceptable range of radius and mass shows that the primary is probably a supergiant of spectral type B0I--B5I. Photometric measurements of its likely counterpart are consistent with the spectral type and luminosity if the distance to the system is between 5.3 kpc and 8.7 kpc. Spectral analysis of the pulsar as a function of orbital phase reveals an evolution of the hydrogen column density suggestive of dense filaments of gas in the downstream wake of the pulsar, with higher levels of absorption seen at orbital phases 0.5--1.0, as well as a variable Fe K$\alpha$ line.

\end{abstract}
 
\section{Introduction}
EXO 1722-363 (IGR J17252-3616) was first discovered by {\em EXOSAT} scan observations in 1984 along with 7 other sources in the Galactic ridge and 14 sources outside the Galactic ridge (Warwick et al. 1988). A pointed {\em Ginga} observation in 1987 October for a total of 80 ks discovered pulsations with a 413.9 $\pm$ 0.2 s period (Tawara et al. 1989). The source exhibited significant variability, with the 6--21 keV flux decreasing from about 2 mCrab to 0.2--0.3 mCrab over 8 hrs. Using an absorbed power-law model, they found $N_{\rm H} = 1.26_{-0.47}^{+0.74} \times 10^{24}$ cm$^{-2}$, suggesting a dense envelope of circumstellar matter surrounding the pulsar. Additional {\em Ginga} observations were performed in 1988 March--April by Takeuchi et al. (1990). Using pulse timing analysis, lower limits on the orbital period and mass of the companion were set to 9 days and 15 $M_{\sun}$, assuming a canonical neutron star mass of 1.4 $M_{\sun}$. Takeuchi et al. (1990) also measured the spectrum but included a high energy cut-off, and found a lower column density $N_{\rm H} = 1.26_{-0.26}^{+0.33} \times 10^{23}$ cm$^{-2}$. 

Using {\em RXTE} Proportional Counter Array (PCA) scans of the Galactic center region from 1999 February to 2003 October, Corbet el al. (2005) measured a 9.741 $\pm$ 0.004 day orbital period through periodic changes in the source flux. By folding the scan observations on this orbital period, eclipses were detected lasting 1.7 $\pm$ 0.1 days (eclipse half-angle $\theta_{e} = 31\fdg8 \pm 1\fdg8$). Corbet et al. (2005) also measured the spectrum using pointed PCA observations from 1998 October--November, and found that the spectrum varies but not dramatically, although the spectrum was sampled over limited ranges of orbital phase. 

EXO 1722-363 has also been observed by {\em INTEGRAL} during surveys of the Galactic plane. Zurita Heras et al. (2006) found that the source is persistent in the 20--60 keV band but is not detected above 60 keV. On a time scale of $\sim$3 days, the 20--60 keV source flux varied by a factor of about 4. Flaring activity was also observed, with the source flux increasing by a factor of 5 and returning to its base level in about a day. Using {\em INTEGRAL}, the pulse period was measured to be 413.7 $\pm$ 0.3 s in 2003 August. Thus, aside from slight ($\sim$1 s) changes, the neutron star pulse period has been stable for over a decade. Phase-resolved spectroscopy did not show any variation of the continuum emission with pulse phase. Using a follow-up {\em XMM-Newton} observation from 2004 March, a 414.8 $\pm$ 0.5 s pulse period was measured. More importantly, however, {\em XMM-Newton} measured the source position to 4\arcsec~accuracy, and an IR counterpart 2MASS J17251139-3616575 is located 1\arcsec~away from the best position. The possible IR counterpart is not detected in the 2MASS survey in the {\em J}-band but appears in the {\it H}-band and {\it K}$_{s}$-band with 11.8 mag and 10.7 mag respectively.

In this work, we use recent and archival {\em RXTE} PCA observations to determine the orbital parameters accurately for \igr~using pulse arrival time analysis. Since the orbital period for this system was previously known, the new observations were timed to provide even coverage in orbital phase, allowing us to obtain firm constraints on the orbital parameters, and especially the eccentricity. The observations are described in \S~2, as well as the pulse arrival time measurements and error analysis. The results of the pulse timing analysis are presented in \S~3.1. We confirm the 9.74 day orbital period and increase its precision by an order of magnitude. The orbital solution is then used to observe orbital modulation of the flux, pulse fraction, and pulse profile in \S~3.2, to constrain the mass and radius of the mass-donating primary in \S~3.3, and to study the orbital modulation of the source spectrum in \S~3.4. We discuss the results of the spectral analysis in \S~4. Finally, in \S~5 we summarize our main results.

\begin{deluxetable*}{cclcccccc} 
\tablenum{1}
\tabletypesize{\scriptsize}
\tablecolumns{9}
\tablewidth{0pt}
\tablecaption{\sc{RXTE Observation Log of EXO 1722-363}} 
\tablehead{
\colhead{ObsID\tablenotemark{a}} &
\colhead{Epoch} &
\colhead{Date} &
\colhead{Exp. Time} &
\colhead{Count Rate\tablenotemark{b}} &
\colhead{Pulse Min.\tablenotemark{c}} &
\colhead{Orb. Phase\tablenotemark{d}} &
\colhead{Spec. Num.\tablenotemark{e}} \\
\colhead{} &
\colhead{} & 
\colhead{(U.T.)} &
\colhead{(ks)} &
\colhead{(Counts/s/PCU)} &
\colhead{} &
\colhead{} &
\colhead{} 
}
\startdata                                            
30142-01-02-01 & 1 & 1998 Oct 23.08--23.25 & 14.80 & 24.77 & 10 & 0.67 & \nodata \\
30142-01-02-00 & 1 & 1998 Oct 27.14--27.31 & 15.14 & 6.30 & 4 & 0.09 & \nodata \\
30142-01-03-00 & 1 & 1998 Oct 31.93--01.10\tablenotemark{f} & 14.22 & 13.13 & 7 & 0.58 & \nodata \\
30142-01-04-00 & 1 & 1998 Nov 05.00--05.17 & 14.16 & 4.18 & 0 & 0.00 & \nodata \\
30142-01-05-00 & 1 & 1998 Nov 09.07--09.25 & 15.84 & 7.89 & 4 & 0.42 & \nodata \\
\hline
80424-01-01-00 & 2 & 2003 Aug 22.21--22.22 & 0.96 & 9.40 & 1 & 0.78 & \nodata \\
80424-01-02-00 & 2 & 2003 Aug 29.58--29.61 & 2.10 & 14.30 & 1 & 0.54 & \nodata \\
\hline
91080-02-04-00 & 3 & 2006 Jan 16.92--17.00 & 4.92 & 4.49 & 0 & 0.00 & 1 \\ 
91080-02-06-00 & 3 & 2006 Jan 17.90--17.99 & 5.04 & 5.21 & 0 & 0.10 & 3 \\ 
91080-02-14-00 & 3 & 2006 Jan 21.84--21.92 & 4.86 & 13.73 & 4 & 0.50 & 11 \\ 
91080-02-08-00 & 3 & 2006 Jan 28.63--28.71 & 4.98 & 7.43 & 1 & 0.20 & 5 \\ 
91080-02-01-00 & 3 & 2006 Feb 04.05--04.13 & 4.74 & 7.94 & 2 & 0.86 & 18 \\ 
91080-02-02-00 & 3 & 2006 Feb 04.52--04.55 & 2.82 & 5.22 & 0 & 0.91 & 19a \\ 
91080-02-01-01 & 3 & 2006 Feb 04.57--04.60 & 2.10 & 5.28 & 0 & 0.91 & 19b \\ 
91080-02-10-** & 3 & 2006 Feb 08.38--08.46 & 4.86 & 25.32 & 4 & 0.30 & 7 \\ 
91080-02-16-00 & 3 & 2006 Feb 11.26--11.30 & 3.54 & 13.50 & 3 & 0.60 & 13 \\ 
91080-02-12-00 & 3 & 2006 Feb 19.18--19.22 & 3.36 & 20.90 & 3 & 0.41 & 9 \\ 
91080-02-15-** & 3 & 2006 Feb 20.56--20.65 & 3.18 & 13.51 & 3 & 0.55 & 12 \\ 
91080-02-18-00 & 3 & 2006 Feb 22.00--22.04 & 3.42 & 13.29 & 3 & 0.70 & 15 \\ 
91080-02-20-00 & 3 & 2006 Feb 23.05--23.09 & 3.18 & 9.82 & 3 & 0.81 & 17 \\ 
91080-02-17-** & 3 & 2006 Mar 03.26--03.40 & 3.18 & 8.76 & 1 & 0.65 & 14 \\ 
91080-02-19-** & 3 & 2006 Mar 04.24--04.39 & 3.24 & 6.90 & 1 & 0.76 & 16 \\ 
91080-02-03-** & 3 & 2006 Mar 06.20--06.29 & 2.88 & 4.18 & 0 & 0.95 & 20 \\ 
91080-02-05-** & 3 & 2006 Mar 07.12--07.20 & 3.30 & 4.12 & 0 & 0.05 & 2 \\ 
91080-02-07-** & 3 & 2006 Mar 08.10--08.19 & 3.42 & 6.05 & 0 & 0.15 & 4 \\ 
91080-02-09-** & 3 & 2006 Mar 09.08--09.17 & 3.48 & 8.27 & 1 & 0.25 & 6 \\ 
91080-02-11-** & 3 & 2006 Mar 10.13--10.22 & 3.42 & 8.68 & 2 & 0.36 & 8 \\ 
91080-02-13-** & 3 & 2006 Mar 11.04--11.13 & 4.02 & 13.05 & 3 & 0.45 & 10 \\ 
\enddata
\tablecomments{Epochs are separated by horizontal lines in the table.}
\tablenotetext{a}{Observation IDs ending with asterisks (**) represent two or more merged observations.} 
\tablenotetext{b}{For 3--24 keV photons after background subtraction.}
\tablenotetext{c}{The number of pulse minima obtained after excluding light segments that are consistent with a constant flux (see text).}
\tablenotetext{d}{The orbital phase at the middle of the observation, inferred from the fit presented in \S~3.1. The phases typically extend about $\pm$0.01 in either direction. }
\tablenotetext{e}{Epoch 3 spectra number labels for Tables 4 \& 5, ordered by orbital phase. Observations 91080-02-02-00 and 91080-02-01-01 were summed for spectral analysis.}
\tablenotetext{f}{This observation ended in November.}
\end{deluxetable*}
\section{Observations}
To find the orbital solution to \igr, we obtained twenty new {\em RXTE} observations (ObsID 91080-02) at carefully chosen intervals during 2006 January to March (hereafter referred to as epoch 3\footnote{Not to be confused with PCA calibration epochs.}), with each observation roughly 3--5 ks in duration. In addition, we used two archival observations: One archival dataset (ObsID 30142-01) contains five observations from 1998 October to November at $\sim$4 day intervals (epoch 1), while the other (ObsID 80424-01) contains two observations from 2003 August of 1 and 2 ks, respectively (epoch 2). A summary of these observations is shown in Table 1. 

\begin{figure}
\centering
\includegraphics[width=3.4in]{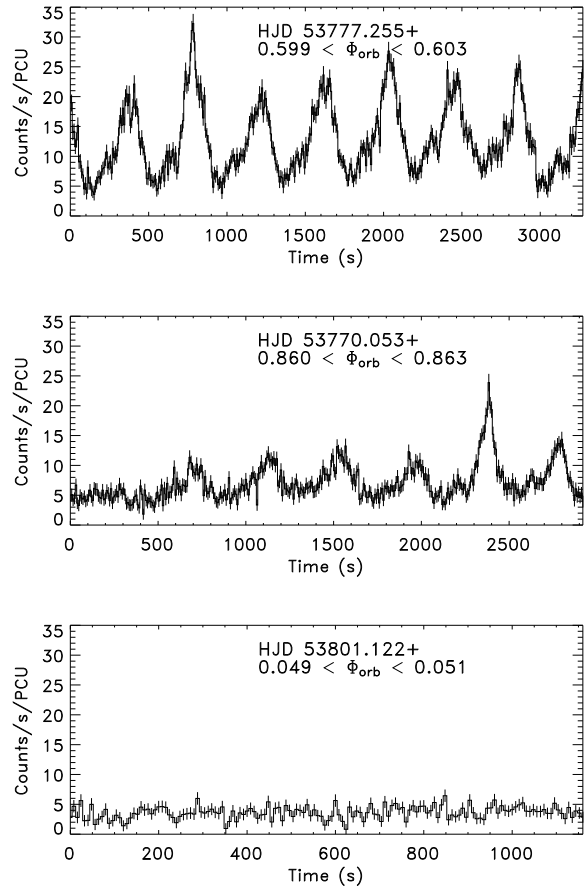}
\caption{Representative light curves at three points in the pulsar's orbit for 3--24 keV photons using 8 s bins. The orbital phases are inferred from the orbital solution in \S~3.1. \label{fig1}}
\end{figure}

Source and background light curves were created with data from the PCA (Jahoda et al. 2006) using standard FTOOLS. Due to the faintness of the source and short observations, HEXTE data were not included in the analysis. The PCA instrument consists of five identical multianode proportional counter units (PCUs), operating in the 2--60 keV range, with an effective area of approximately 6500 cm$^{2}$ and a 1$\degr$ FWHM field of view. Photon energies for pulse timing analysis were restricted to 3--24 keV, and the light curves were binned by 1 s. Data accumulated in the Good Xenon data mode was used when available. The Faint\footnote{See http://heasarc.nasa.gov/docs/xte/pca\_news.html.} background model was used for the all epochs. The arrival times of individual events were reduced to the solar system barycenter using the Jet Propulsion Laboratory DE-200 ephemeris and the FTOOL {\tt faxbary}. Figure 1 shows representative epoch 3 light curves at three different points in the orbit. Periodic decreases in source flux, called dips, are observed in sufficiently inclined systems that have an accretion disk. These periodic decreases are thought to be caused by a bulge of material where the accreted matter impacts the disk, obscuring the X-ray source. Such dips typically last for several hours. By examining the low-energy light curve in detail (where photoelectric absorption has the largest effect), in addition to the hardness ratio, we do not find any evidence for dips in \igr.

\subsection{Arrival Times}
\begin{figure}
\centering
\includegraphics[width=3.35in]{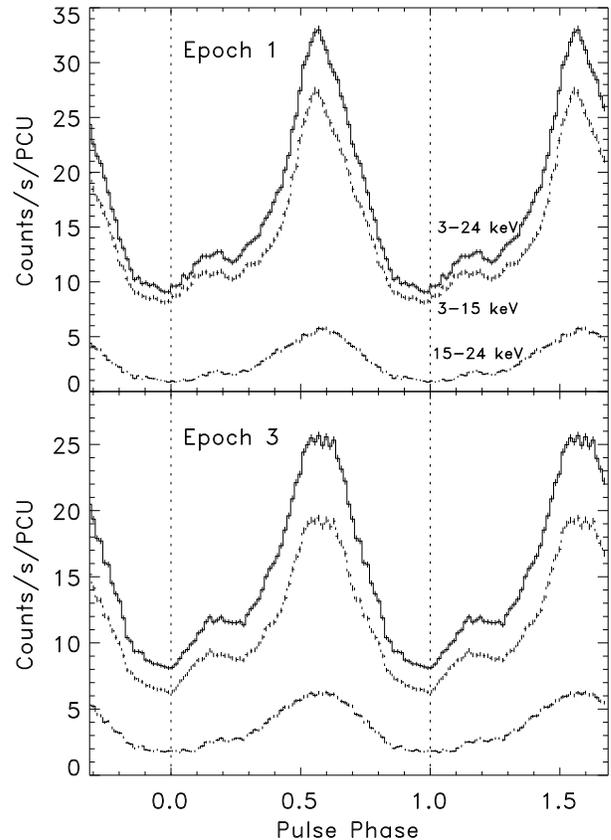}
\caption{Final pulse templates for epochs 1 and 3 that were used to find the pulse arrival times via cross-correlation, for 3--24 keV photons ({\em solid} line), 3--15 keV photons ({\em dotted} line), and 15--24 keV photons ({\em dashed} line). The vertical dotted lines represent the phase corresponding to the pulse arrival times (Table 2). The error bars are 1$\sigma$ statistical uncertainties.\label{pprofile}}
\end{figure}

Pulse arrival times were measured in a multi-step iterative process, similar to the analysis in Thompson et al. (2006). First, each individual observation was divided into $\sim$1.5 ks ``light segments'' ($\sim$3--4 pulse cycles), and each segment was folded modulo 414.0 s to improve statistics. Although the true pulse period may differ slightly from this value, we expect smearing of the pulsation to be minimal, leading to systematic errors in the measured pulse arrival times of $\la$ 1 s. \igr~is known to be an eclipsing system (Corbet et al. 2005). Zurita Heras et al. (2006) found that the {\em INTEGRAL} ISGRI flux drops to zero during the eclipse. The PCA flux, on the other hand, does not drop to zero during eclipse, probably because of Galactic ridge emission or possible weak contaminating sources within the 1$\degr$ field of view.  To limit the potential for including spurious pulse minima when the pulsar is obscured by the high-mass companion, each observation was fit with a constant flux and $\chi^{2}$ was calculated. Based on the null hypothesis that the light segments are consistent with a constant flux, pulse minima were only measured if the null hypothesis was rejected at the 1\% level of significance. While this may exclude data when the pulsar is out of eclipse, we prefer to be conservative and potentially exclude usable segments rather than include segments in eclipse. With the orbital solution obtained in \S~3.1, we find that all excluded light curves are within $\vline \Phi_{\rm orb} \vline < 0.15$, where mid-eclipse is defined to be orbital phase 0.0. The eclipse duration implies that pulsations should be observable at phase 0.1. Therefore, although all of the excluded observations occurred near the eclipse, they did not necessarily take place during the eclipse. The light segments that were not consistent with a constant were then manually aligned to create a preliminary pulse template. Each light segment was cross correlated with the preliminary pulse template, and the cross correlation lag times were then used to create a more refined pulse template. This process was repeated until the pulse arrival times and the pulse template no longer changed, and was performed on the epoch 1 and epoch 3 data separately. The epoch 3 pulse template was used to find the epoch 2 pulse minima. In total, 61 pulse arrival times were obtained. They are presented with their associated errors in Table 2. The final pulse templates for epochs 1 and 3 are shown in Figure \ref{pprofile}. It is interesting to note that the epoch 1 pulse profile is more peaked at $\phi \approx$ 0.5--0.6 than the epoch 3 pulse profile.


\begin{deluxetable}{lc|lc} 
\tablenum{2}
\tabletypesize{\scriptsize}
\tablecolumns{4}
\tablewidth{0pt}
\tablecaption{\sc{Pulse Arrival Times}} 
\tablehead{
\colhead{Arrival Time\tablenotemark{a}} &
\colhead{Statistical/Total} \vline &
\colhead{Arrival Time\tablenotemark{a}} &
\colhead{Statistical/Total} \\
\colhead{} & 
\colhead{Uncertainty\tablenotemark{b}} \vline &
\colhead{} &
\colhead{Uncertainty\tablenotemark{b}} \\
\colhead{(s)} &
\colhead{(s)} \vline & 
\colhead{(s)} &
\colhead{(s)}
}
\startdata
0 & 2.3/5.0 & 229359623 & 3.1/6.7 \\       
823 & 1.6/4.8 & 229910149 & 2.6/6.5 \\
1650 & 1.8/4.8 & 229910977 & 1.6/6.1 \\
5798 & 1.5/4.7 & 230282137 & 0.5/5.9 \\
6627 & 1.5/4.7 & 230283372 & 0.5/5.9 \\
7455 & 1.7/4.8 & 230284614 & 0.5/5.9 \\
8287 & 2.1/4.9 & 230287911 & 1.3/6.1 \\
11604 & 0.9/4.6 & 230531238 & 2.8/6.5 \\
12439 & 2.0/4.9 & 230532480 & 0.3/5.9 \\
13674 & 1.0/4.6 & 230533308 & 1.1/6.0 \\
351102 & 27.6/30.0 & 231216181 & 0.5/5.9 \\
353169 & 11.3/12.2 & 231217016 & 1.0/9.2 \\
356875 & 3.0/5.4 & 231218242 & 0.5/5.9 \\
363493 & 3.0/5.4 & 231334948 & 2.5/6.2 \\
765590 & 1.4/4.7 & 231340752 & 0.9/7.2 \\
771385 & 1.7/4.8 & 231341580 & 0.7/6.8 \\
772214 & 2.8/5.2 & 231459595 & 1.5/6.4 \\
773458 & 2.7/5.2 & 231460421 & 1.4/6.4 \\
777171 & 1.5/4.7 & 231461665 & 1.4/6.1 \\
778004 & 2.8/5.3 & 231550305 & 3.1/6.7 \\
778837 & 1.0/4.6 & 231551114 & 1.7/6.2 \\
1468308 & 3.1/5.5 & 231552374  & 1.0/6.0 \\ 
1470385 & 2.3/5.0 & 232259983 & 6.0/8.4 \\
1474508 & 2.2/5.0 & 232344878 & 2.6/6.5 \\
1475756 & 2.6/5.2 & 232762873 & 1.9/6.2 \\
152421254 & 5.2/7.2 & 232853859 & 2.7/6.5 \\
153058868 & 3.6/6.2 & 232859646 & 3.6/6.9 \\
228766823 & 0.8/6.0 & 232932443 & 1.6/6.1 \\
228770532 & 2.1/6.3 & 232933281 & 4.4/7.4 \\
228771378 & 1.2/6.0 & 232938665 & 0.8/6.0 \\
228772626 & 0.3/5.9 & \nodata & \nodata\\
\enddata
\tablenotetext{a}{Arrival times begin at HJD  51109.07.} 
\tablenotetext{b}{A systematic uncertainty of 4.5 (5.9) s was added to each epoch 1 (3) arrival time. The epoch 2 data consist of only two minima, so no systematic deviation could be obtained. We chose to apply a value of 5 s.}
\end{deluxetable}

\subsection{Error Analysis}
Statistical errors on the pulse arrival times were calculated using Monte Carlo simulations. The PCA background errors were estimated from the unmodeled background variations shown in Fig. 29 of Jahoda et al. (2006), ranging from $\sim$3--8\% over 3--24 keV with 16 s binning. The background and errors were interpolated to 1 s resolution. We refer the reader to Thompson et al. (2006) for an explanation of the simulations. 

\begin{figure}
\centering
\includegraphics[width=2.75in]{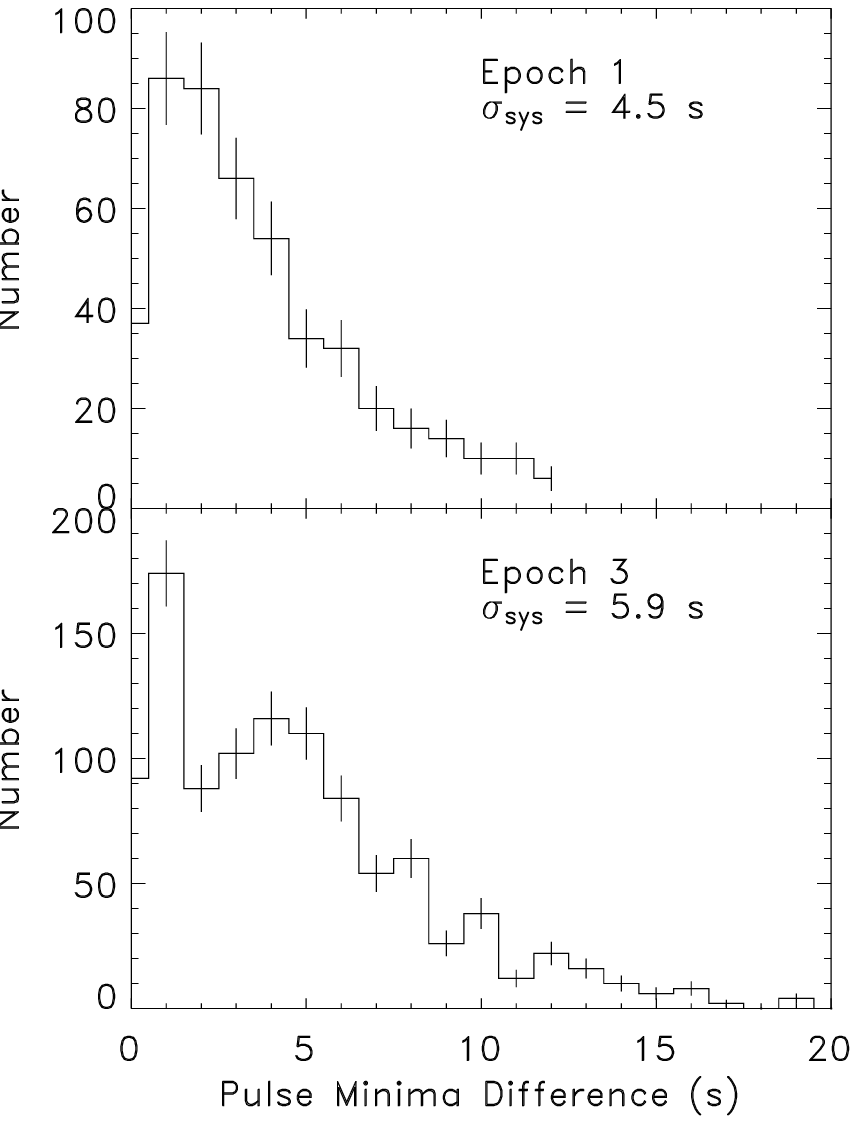}
\caption{Histogram of the magnitude of the difference in pulse minima versus the final template created by using all light segments as pulse templates for all other folded light curves for epochs 1 and 3. \label{syserrhist}}
\end{figure}

In addition to statistical errors, systematic uncertainties in the pulse arrival times can be caused by changes in the pulse profile due to varying flux levels, or varying levels of local absorption. To estimate these errors, we aligned each folded light segment with the pulse template using the measured pulse arrival times, and each aligned and folded light segment was then used as a trial pulse template for all of the other folded light segments. Figure \ref{syserrhist} shows the resulting histograms of the magnitude of the difference in the measured pulse minima versus the final templates for epochs 1 and 3. We find a systematic deviation of 4.5 s for the epoch 1 data, and 5.9 s for the epoch 3 data. Epoch 2 consists of only two observations, so no systematic deviation can be obtained. We therefore chose to apply a value of 5 s to these data. The systematic errors were added to the statistical errors in quadrature, which may slightly overestimate the total errors because the statistical and systematic errors are likely correlated.

\section{Results}
\subsection{Pulse Timing Analysis}
\begin{figure*}
\centering
\includegraphics[width=4in]{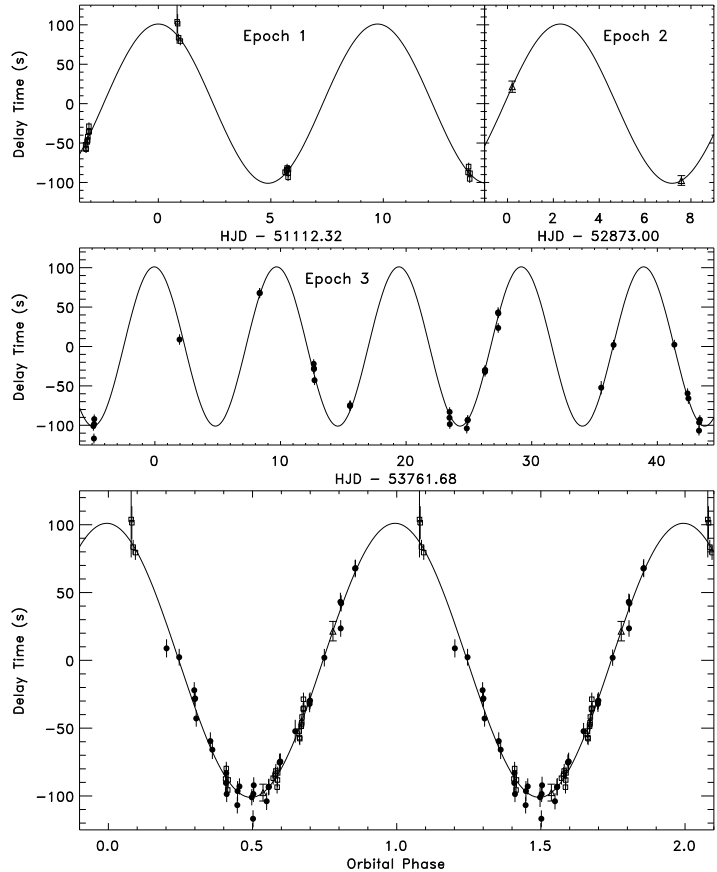}
\caption{Orbital solution to EXO 1722-363. {\em Top} and {\em middle} panels: Plots of the fits to each epoch. The epoch 1 (3) data is plotted with open squares (filled circles). The epoch 2 data is plotted with open triangles and have hats on the error bars. {\em Bottom} panel: All epochs plotted modulo the orbital period for two cycles.\label{fit}}
\end{figure*}

The pulse arrival times were fitted with a seven parameter model of the orbit and pulse period evolution. The arrival time of the $n$th pulse is given by
\begin{equation} 
t_{n}=t_{0}+nP_{0}+\frac{n^{2}}{2}P_{0}  \mbox{\.{\em P}} + a_{x}\sin{i}\cos{\left[\frac{2\pi \left( t_{n}-{\rm T0}\right)}{P_{\rm orb}}\right]},
\end{equation}
where $P_{0}$ is the pulse period at time $t_{0}$, $\mbox{\.{\em P}}$ is the pulse period derivative and is assumed to be constant, $a_{x}\sin{i}$ is the projected semi-major axis of the orbit, $P_{\rm orb}$ is the orbital period of the system, and T0 is a reference time corresponding to mid-eclipse for a circular orbit. The pulse number $n$ is given by the nearest integer to
\begin{equation}
n=\frac{t_{n}-t_{0}}{\langle P \rangle}=\frac{t_{n}-t_{0}}{P_{0}+0.5 \mbox{\.{\em P}}(t_{n}-t_{0})}.
\end{equation}
Because the epoch 3 data has the greatest phase coverage, we first fit these pulse minima alone, yielding an orbital solution with $a_{x}\sin{i} = 100 \pm 4$ lt-s and $P_{\rm orb} = 9.78 \pm 0.04$ days (90\% confidence). Based on the preliminary epoch 3 fit, we attempted to pulse-connect the fit to the epoch 2 data from 2.4 yrs earlier, using the assumption of a single \mbox{\.{\em P}} between the datasets. A convincing connection was made which requires no pulse number ambiguity over more than 180,000 cycles. The pulse period derivative expressed as a fractional rate is $\mbox{\.{\em P}}/P=-7.6 \times 10^{-4}$ yr$^{-1}$. The resulting projected semi-major axis remained the same, but the orbital period was refined to $9.741^{+0.005}_{-0.003}$ days. With the additional constraints on the orbital period from this connection, we were able to connect the data through orbits to the epoch 1 data from 7.3 yrs before epoch 3 (272 orbital cycles). For this fit, however, the pulses could not be connected with a single \mbox{\.{\em P}}, and so the epoch 1 pulse period and pulse period derivative were allowed to vary independently. The precision of the projected semi-major axis increased slightly with $a_{x}\sin{i} = 101 \pm 3$ lt-s, and the orbital period was further refined to $9.7403 \pm 0.0004$ days. Furthermore, the mass function was found to be $f_{x}(M)=11.7 \pm 1.2$ $M_{\sun}$, and we define an epoch of mid-eclipse at HJD 53761.68 $\pm$ 0.04. The results of the fits are presented in Table 3, and the pulse arrival delays are plotted in Figure \ref{fit}. The error bars in the figures represent the 68\% confidence level, and the secular change in the pulse periods has been removed so that the modulation is purely due to orbital motion. By including an additional term in equation (1) representing the first-order term in a Taylor series expansion in the eccentricity (Levine et al. 2004), the eccentricity is 0.08$^{+0.11}_{-0.08}$, the projected semi-major axis is 96 $\pm$ 7 lt-s, and the orbital period is 9.7400$^{+0.0006}_{-0.0007}$ days. However, the fit does not improve significantly, and with additional model parameters the value of $\chi^{2}_{\nu}$ increases slightly from 1.40 to 1.42. The \igr~orbit is therefore consistent with circular, and $e <0.19$ with 90\% confidence.
\begin{deluxetable*}{lccccc} 
\tablenum{3}
\tabletypesize{\scriptsize}
\tablecolumns{6}
\tablewidth{0pt}
\tablecaption{\sc{Orbital Solution to EXO 1722-363}\label{fits}} 
\tablehead{
\colhead{Parameter} & 
\colhead{(units)} &
\colhead{Epoch 3} & 
\colhead{Epochs 2 \& 3} & 
\multicolumn{2}{c}{Epochs 1--3\tablenotemark{a}}
}
\startdata
$P_{\rm pulse}$\tablenotemark{b} & (s) & 413.895 $\pm$ 0.002 & 413.8933$^{+0.0002}_{-0.0006}$ & 413.851 $\pm$ 0.004 & 413.8933$^{+0.0002}_{-0.0006}$  \\
$\mbox{\.{\em P}}_{\rm pulse}$ & ($\times 10^{-9}$ s s$^{-1}$) & -10.66 $\pm$ 0.99 & -9.98 $\pm$ 0.004 & 0 $\pm$ 6 & -9.98$^{+0.015}_{-0.004}$ \\
$a_{x}\sin{i}$ & (lt-s) & 100 $\pm$ 4 & 100 $\pm$ 4 & \multicolumn{2}{c}{101 $\pm$ 3} \\
$P_{\rm orb}$ & (d) & 9.78 $\pm$ 0.04 & 9.741$^{+0.005}_{-0.003}$  & \multicolumn{2}{c}{9.7403 $\pm$ 0.0004} \\
$f_{x}(M)$ & $({\rm M}_{\sun})$ & 11.2 $\pm$ 1.3 & 11.3 $\pm$ 1.4 & \multicolumn{2}{c}{11.7 $\pm$ 1.2} \\ 
T0\tablenotemark{c} & (HJD) & 53761.60 $\pm$ 0.09 & 53761.68 $\pm$ 0.04 & \multicolumn{2}{c}{53761.68 $\pm$ 0.04} \\
$\chi^2_{\nu}$ (dof) & & 1.46 (28) & 1.47 (30) & \multicolumn{2}{c}{1.40 (53)} \\
\enddata
\tablecomments{All errors are quoted at the 90\% confidence level for a single parameter.}
\tablenotetext{a}{For this fit, $P_{\rm pulse}$ and $\mbox{\.{\em P}}_{\rm pulse}$ were varied independently for epoch 1 ({\em left}) and epochs 2 \& 3 ({\em right}).} 
\tablenotetext{b}{Pulse periods at $t_{0} = {\rm HJD}~51112.18662$ for the fits to epochs 1, and $t_{0} = {\rm HJD}~53761.73126$ for the fits to epochs 2 \& 3.} 
\tablenotetext{c}{T0 is defined to the epoch of mid-eclipse at $\Phi_{\rm orb}=0$.}
\end{deluxetable*}

\begin{figure}
\centering
\includegraphics[width=3.3in]{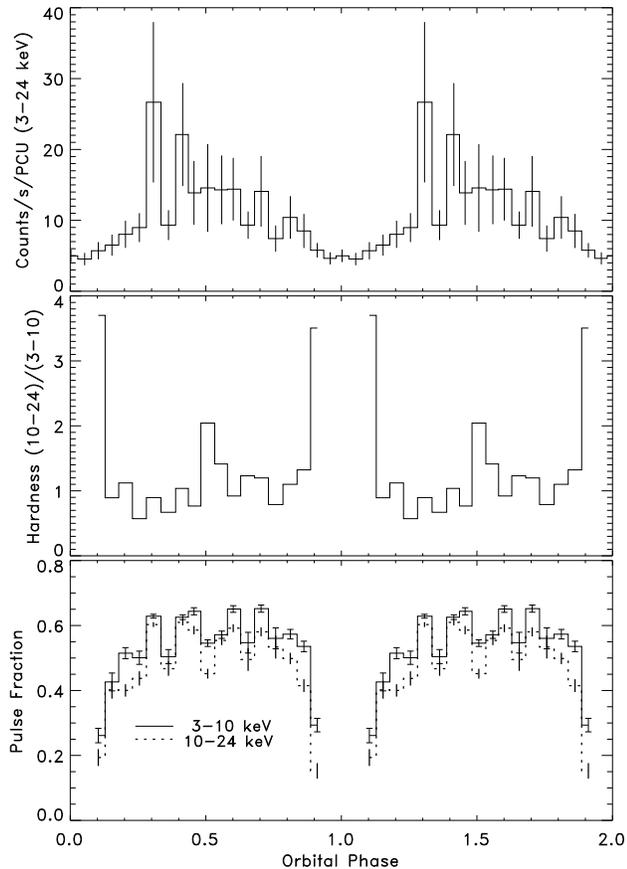}
\caption{Orbital modulation of the mean flux ({\em top panel}), the hardness ratio ({\em middle panel}), and the pulse fraction ({\em bottom panel}) during epoch 3. The error bars for the top panel represent 1$\sigma$ standard deviations (the uncertainty in the averages are less than 1\%), and those of the bottom panel represent 1$\sigma$ statistical uncertainties. The hardness ratio and pulse fraction are corrected for the Galactic ridge emission. The pulse fraction is shown in two energy bands; the 3-10 keV data ({\em solid line}) have hats on the error bars, and the 10--24 keV data ({\em dotted line}) do not have hats on the error bars.\label{orbmod}}
\end{figure}
\begin{figure*}
\centering
\includegraphics[width=4in]{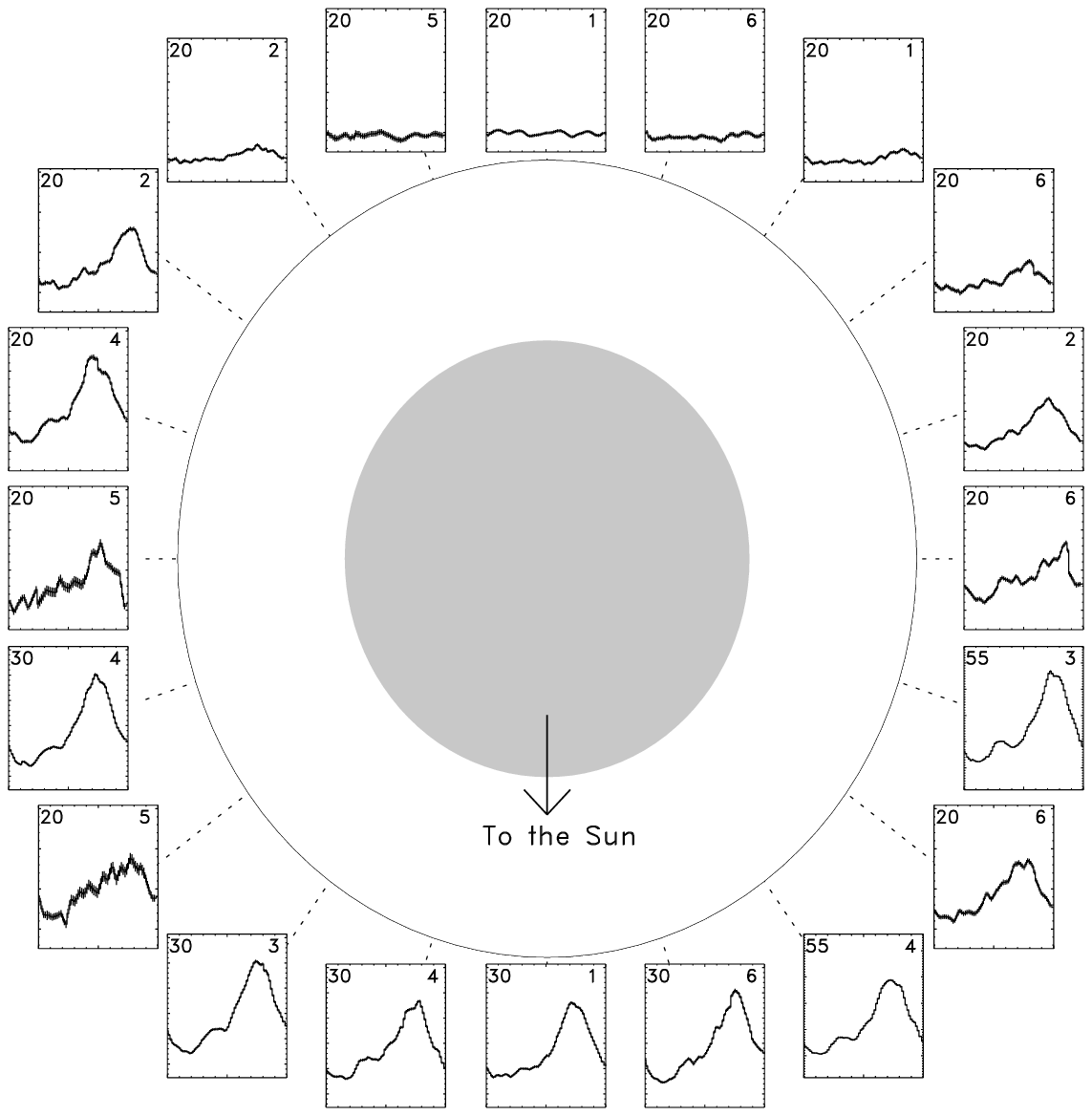}
\caption{Orbital modulation of the epoch 3 average pulse profile from 3--24 keV. The axis labels are omitted for clarity: The $x$-axes is pulse phase from 0.0 to 1.0 and the $y$-axes is counts s$^{-1}$ PCU$^{-1}$ from 3 to the number in the upper left corner of each panel. The orbit from which the pulse profile came is listed in the upper right corner or each panel. Pulsations cannot be seen during eclipse, and can only be seen weakly at phases 0.1 and 0.9. The system is viewed from below, and the circle represents the orbit of the pulsar. The size of the projected stellar radius relative to the size of the projected orbit is based on the Corbet et al. (2005) eclipse time measurement. \label{ppfig} }
\end{figure*}

\subsection{Orbital Modulation of Flux, Pulse Fraction, and Pulse Profile}
Using the orbital solution, we searched for orbital modulation of the mean flux, the hardness ratio, and the pulse fraction using 20 orbital phase bins, shown in Figure \ref{orbmod}. This figure clearly shows the orbital modulation of the flux, and the general shape is consistent with the {\em RXTE} ASM light curve presented by Corbet et al. (2005, Fig. 10) and the {\em INTEGRAL} ISGRI light curve of Zurita Heras et al. (2006, Fig. 9), showing asymmetrical modulation of the orbital flux with a slightly higher average flux at orbital phases when the pulsar is moving towards the solar system barycenter than when the pulsar is moving away. The hardness ratio and pulse fraction are corrected for the Galactic ridge emission, which was measured from three observations occurring during eclipse (see below). The hardness ratio (using the PCA count rates in the 10--24 keV and 3--10 keV bands) is fairly constant except for the observations occurring near the eclipse, and at orbital phase 0.5. The pulse fraction for $0.15 < \Phi_{\rm orb} < 0.85$ varies between 40\% and 65\% in both energy bands, but decreases outside this range and is consistent with zero during eclipse. We speculate that the decrease in pulse fraction near eclipse ingress and egress may be due to smearing of the pulsations from Thomson scattering along the line of sight. Figure \ref{ppfig} shows modulation of the average pulse profile as a function of the position of the pulsar in the orbit, created by folding all data in each phase bin, and accounting for slight changes in the pulse period and for the pulsar's orbital position. 

\subsection{Properties of the Mass-Donating Companion}
\begin{figure}
\centering
\includegraphics[width=3.3in]{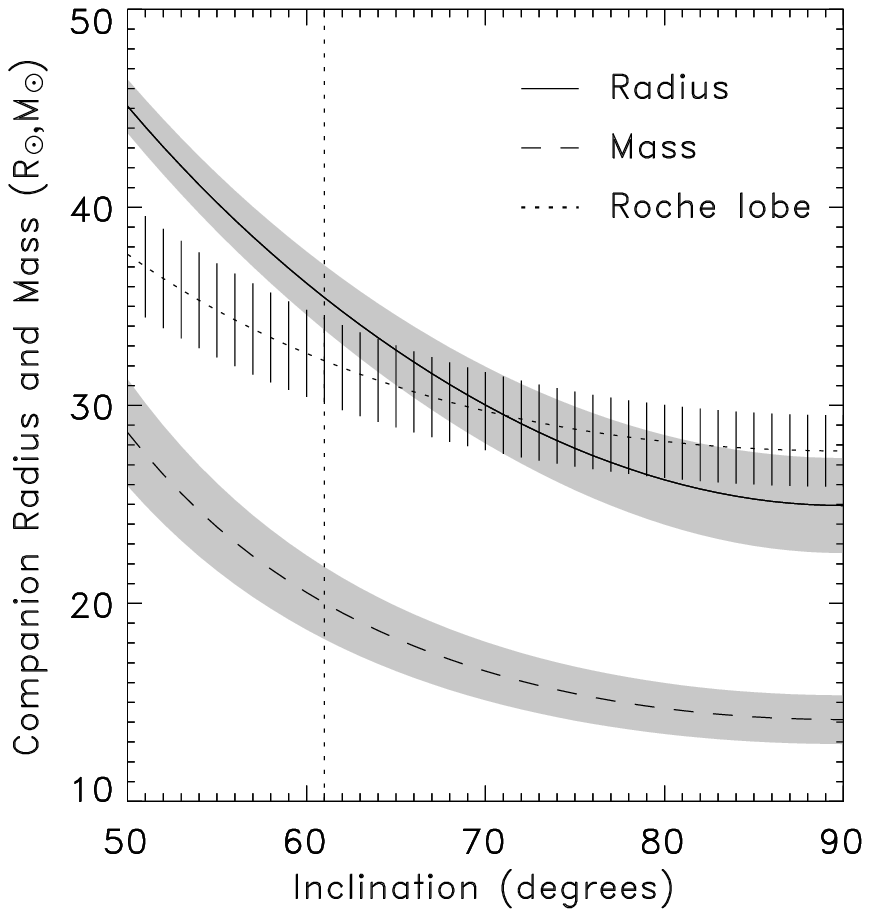}
\caption{Inclination versus companion radius and mass, plus the size of the Roche lobe, using the best-fit orbital parameters from \S~3.1, the Corbet et al. (2005) eclipse half-angle measurement, and assuming a neutron star mass of 1.4 $M_{\sun}$. The $y$-axis scale applies to both radius and mass although the units differ. The radius and mass curves have shaded error contours, and the Roche lobe curve has error bars. The vertical dotted line at $i=61\degr$ represents the lower limit to the inclination at the 99\% confidence level. \label{incl}}
\end{figure}

Figure \ref{incl} shows the inferred companion radius and mass, as well as the Roche lobe radius (Eggleton 1983), as a function of the binary inclination. The curves were derived using our orbital solution, the Corbet et al. (2005) eclipse half-angle measurement, and a neutron star mass of 1.4 $M_{\sun}$. As shown in the figure, the radius of the companion ($R_{c}$) is greater than about 22 $R_{\sun}$, and at the limit of a zero mass neutron star $R_{c} > 21$ $R_{\sun}$. The figure also allows us to constrain the inclination assuming the primary underfills its Roche lobe. Since the uncertainties in the Roche lobe and primary radii are correlated through their interdependence on $a_{x} \sin{i}$, simulations were used to find that $i > 61\degr$ at the 99\% confidence level. With this firm constraint on the inclination, the radius of the primary must be less than about 37 $R_{\sun}$ and its mass must be less than about 22 $M_{\sun}$. 

The radius and mass constraints are consistent with a supergiant star with a spectral type between B0 and B5 (Cox 2000). By combining photometric measurements by the Two Micron All-Sky Survey (2MASS) and the Deep Near-Infrared Survey of the Southern Sky (DENIS) in the {\em J-} and {\em K}-bands, comparing the colors that are obtained with empirical $(J - K)_{0}$ values for B0I--B5I stars, and applying extinction relationships ($A_{\lambda}/A_{V}$) from Cardelli et al. (1989), we find that the visual extinction is $A_{V} = 20.1 \pm 0.6$. The absolute visual magnitude for B0--B5 supergiants is $M_{V} \approx -6.5$ to $-6.2$ (Cox 2000), and the inferred spectral type is only consistent with the {\em J} and {\em K} apparent magnitudes if the distance to the system is between 5.3 kpc and 8.7 kpc. 
\begin{deluxetable}{lc} 
\tablenum{4}
\tabletypesize{\scriptsize}
\tablecolumns{2}
\tablewidth{0pt}
\tablecaption{\sc{Eclipse Spectrum for EXO 1722-363}\label{bkgspec}} 
\tablehead{
\colhead{Parameter} & 
\colhead{Value}
}
\startdata
$N_{\rm H}$ ($\times 10^{22}$ cm$^{-2}$) & 7.1$^{+2.2}_{-2.3}$ \\
Raymond-Smith $kT$ (keV) & 2.2$^{+0.5}_{-0.3}$ \\
\hspace{0.1in} Normalization ($\times 10^{-2}$) & $9.0^{+4.2}_{-3.5}$ \\
Power law (PL) Index & 1.8$^{+0.1}_{-0.3}$ \\
\hspace{0.1in} Normalization ($\times 10^{-3}$) & $6.2^{+5.6}_{-3.1}$ \\
\enddata
\tablecomments{All errors are quoted at the 90\% confidence level for a single parameter. Spectra numbers 1, 2, \& 20 were fit simultaneously from 3--24 keV; $\chi^2/\nu=118.2/136=0.87$. The abundances of the Raymond-Smith plasma model were fixed at solar values. Model normalizations follow the definitions used in XSPEC (Arnaud 1996).}
\end{deluxetable}

\subsection{Spectral Analysis}
\begin{figure}
\centering
\includegraphics[width=3.3in]{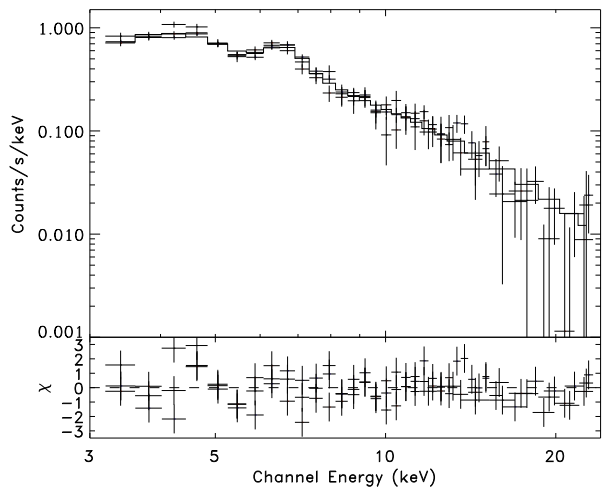}
\caption{EXO 1722-363 spectrum and residuals for the three observations (spectra 1, 2, \& 20) that occurred during eclipse. Emission is assumed to be dominated by the GRE. The best-fit model was fixed for the fits to spectra numbers 3--19. \label{bkgfig}}
\end{figure}

The epoch 3 PCA observations have excellent orbital phase coverage, allowing us to observe changes in the source spectrum as the pulsar travels through its orbit. For spectral analysis, all datasets occurring within a single 4 hr period were summed (see labels in Table 1). Although PCUs 0 and 2 were on throughout the observations, the loss of the propane layer in PCU 0 has resulted in a higher background level. Since EXO 1722-363 is a relatively weak source, we only used data from PCU 2 to minimize concerns with background subtraction. Because of the 1${\degr}$ field of view of the PCA and EXO 1722-363's position in the Galactic plane ($b=-0\fdg 35$), the spectrum consists of emission from the pulsar plus diffuse emission from the Galactic ridge. The Galactic ridge emission (GRE) was modeled following Valinia \& Marshall (1998), using an absorbed Raymond-Smith plasma with temperature $\sim$2--3 keV plus a power-law component with a photon index of $\sim$1.8. Fortunately, due to the presence of eclipses in EXO 1722-363, the GRE can be modeled by simultaneously fitting all observations occurring during eclipse. As pointed out by Corbet et al. (2005), the epoch 1 {\em RXTE} observation 30142-01-02-00 shows an apparent egress from eclipse (see their Fig. 5). Applying our orbital solution and assuming a symmetric eclipse, we obtained a 1.68 $\pm$ 0.06 day eclipse duration, consistent with 1.7 $\pm$ 0.1 days by Corbet et al. (2005), but inconsistent with the $\sim$1.3 day measurement by Zurita Heras et al. (2006). Note that our uncertainty is partly due to a lack of ability to define precisely when pulsations become apparent. Taking the lower limit of the Corbet et al. value at 1.6 days translates into $0.92 < \Phi_{\rm orb} < 0.08$. From this, we find that three of the epoch 3 observations took place during eclipse (spectra numbers 1, 2, \& 20). The GRE parameter values were then obtained by simultaneously fitting the spectra using XSPEC v11.3.2, the results of which are presented in Table 4. The model fit and residuals are shown in Figure \ref{bkgfig}. This model also accounts for any X-ray emission from the supergiant primary and scattered emission from the pulsar. 

\begin{figure}
\centering
\includegraphics[width=3.3in]{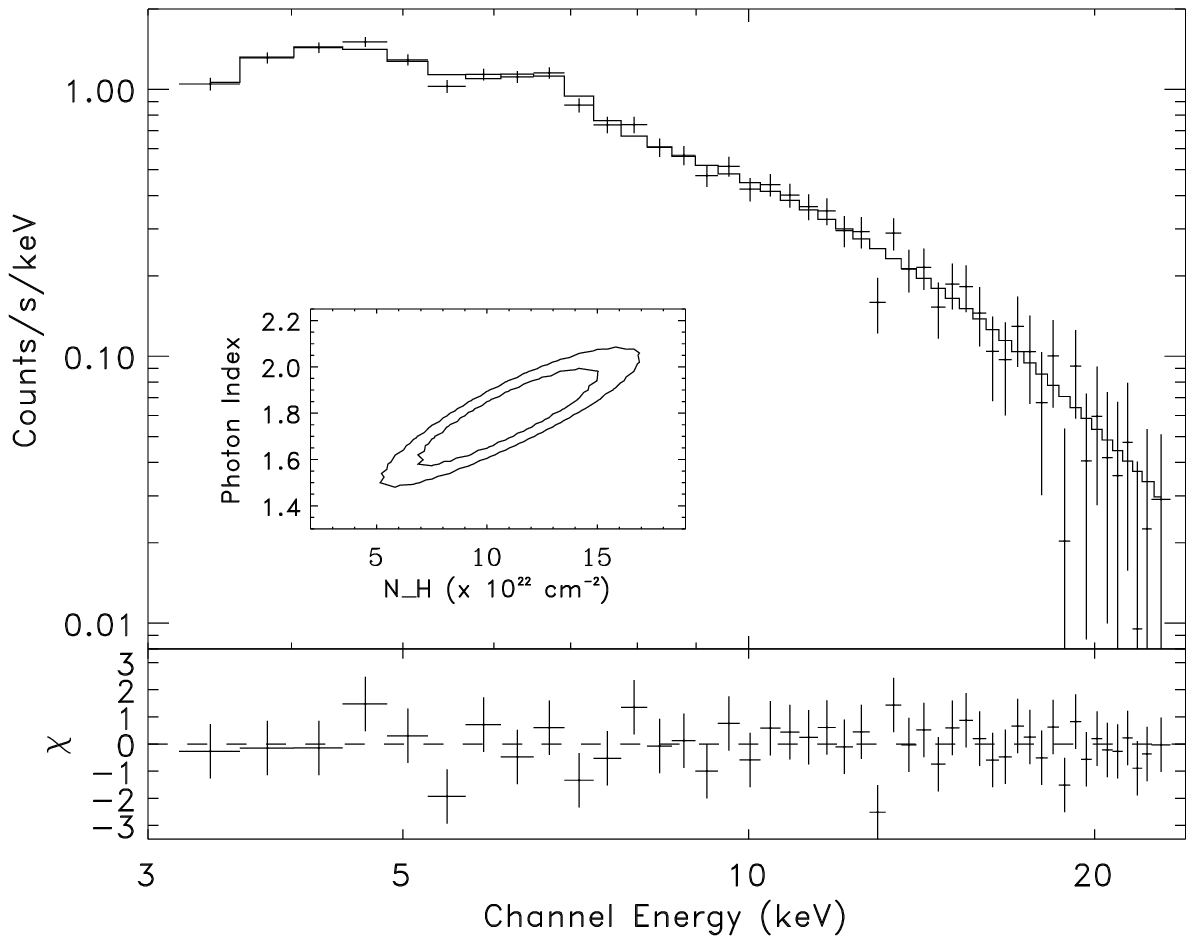}
\caption{EXO 1722-363 spectrum number 8 and residuals. {\em Inset:} Contour plot of $N_{\rm H}$ versus photon index, showing the 68\% and 90\% confidence intervals ($\Delta \chi^2 = 2.3$  and $\Delta \chi^2 = 4.6$, respectively). \label{srcfig}}
\end{figure}

With the GRE model parameters fixed at the best-fit values, the 17 remaining spectra were fit with an absorbed power law with a high energy cut-off plus a narrow Gaussian fixed at 6.4 keV to model iron line emission. Zurita Heras et al. (2006) found that phase-resolved spectroscopy did not show any variation of the continuum spectral shape with pulse phase, so pulse phase-resolved spectral analysis is unnecessary. Limited statistics made a high energy cut-off unnecessary for 11 of the 17 spectra; however, we believe that the cut-off is probably still present because it is a common characteristic of pulsar spectra (Mihara 1995). We therefore chose to include a high energy cut-off in all spectra, but allowed the cut-off and folding energies to vary only if an {\em F}-test showed that there is less than a 5\% probability that the improvement to the fit due to the cut-off occurred by chance. With this criterion, spectra numbers 7, 9, 11--13, \& 16 had free values of $E_{\rm cut}$ and $E_{\rm fold}$. In all other cases, the cut-off and folding energies were fixed to the best-fit value found by fitting all spectra simultaneously with a single cut-off and folding energy. The best-fit cut-off energy was $16.2^{+0.8}_{-1.2}$ keV and the best-fit folding energy was $22.0^{+5.8}_{-4.7}$ keV. The spectral analysis results are shown in Table 5. As an example, we show the fit and residuals to spectrum number 8 (near $\Phi_{\rm orb} = 0.36$) in Figure \ref{srcfig}. Also shown (see inset) are the 68\% and 90\% confidence intervals ($\Delta \chi^2 = 2.3$  and $\Delta \chi^2 = 4.6$, respectively) for $N_{\rm H}$ versus photon index. Note that the two parameter confidence interval implies uncertainties in $N_{\rm H}$ (photon index) that are about 20\% (5\%) larger than those quoted in Table 5. After subtracting the GRE flux, we find the unabsorbed X-ray flux to be in the range 0.7--4.8 $\times 10^{-10}$ ergs cm$^{-2}$ s$^{-1}$ from 2--24 keV. With the distance constraints from above, the unabsorbed X-ray luminosity is $L_{x} \sim$ 0.2--4.4 $\times 10^{36}$ ergs s$^{-1}$ over the same range.
\begin{deluxetable*}{ccccccccc} 
\tablenum{5}
\tabletypesize{\scriptsize}
\tablecolumns{9}
\tablewidth{0pt}
\tablecaption{\sc{Epoch 3 Spectral Fits to EXO 1722-363}} 
\tablehead{
\colhead{Number} &
\colhead{Orb. Phase\tablenotemark{a}} &
\colhead{$N_{\rm H}$} &
\colhead{Photon Index} &
\colhead{PL Norm.} &
\colhead{Cut-off Energy\tablenotemark{b}} &
\colhead{Folding Energy\tablenotemark{b}} &
\colhead{EW\tablenotemark{c}} &
\colhead{$\chi^{2}_{\nu}$} \\
\colhead{} &
\colhead{} & 
\colhead{($\times 10^{22}$ cm$^{-2}$)} &
\colhead{} &
\colhead{($\times 10^{-3}$)} &
\colhead{(keV)} &
\colhead{(keV)} &
\colhead{(eV)} &
\colhead{(dof)} 
}
\startdata                                     
3 & 0.10 & $86^{+87}_{-61}$ & $0.8^{+1.8}_{-1.3}$ & $0.6^{+34.2}_{-0.5}$ & 16.2 & 22.0 & $<4840$ & 0.70 (43) \\
4 & 0.15 & $20^{+16}_{-12}$ & $2.0^{+0.7}_{-0.3}$ & $12.2^{+58.1}_{-6.8}$ & 16.2 & 22.0 & $<335$ & 0.65 (43) \\
5 & 0.20 & $36^{+11}_{-12}$ & $1.7^{+0.4}_{-0.3}$ & $14.3^{+22.1}_{-9.3}$ & 16.2 & 22.0 & $<170$ & 0.67 (43) \\
6 & 0.25 & $8^{+5}_{-4}$ & $1.8^{+0.3}_{-0.2}$ & $13.0^{+12.2}_{-6.0}$ & 16.2 & 22.0 & $<200$ & 1.14 (43) \\
7 & 0.30 & $11 \pm 2$ & $1.2 \pm 0.1$ & $18.6^{+3.8}_{-3.3}$ & 16.6$^{+1.2}_{-1.7}$ & 19.0$^{+10.0}_{-5.9}$ &  174$^{+92}_{-54}$  & 0.68 (41) \\
8 & 0.36 & $11 \pm 5$ & $1.8 \pm 0.2$ & $15.3^{+11.6}_{-6.9}$ & 16.2 & 22.0 & $<133$ & 0.71 (43) \\
9 & 0.41 & $10 \pm 2$ & $0.9 \pm 0.1$ & $8.2^{+2.6}_{-2.1}$ & 16.5$^{+1.9}_{-1.7}$ & 17.3$^{+10.4}_{-7.1}$ & 352$^{+212}_{-144}$ &  0.61 (41) \\
10 & 0.45 & $10 \pm 3$ & $1.4 \pm 0.1$ & $12.9^{+2.3}_{-3.9}$ & 16.2 & 22.0 & $158^{+75}_{-116}$ & 0.93 (43) \\
11 & 0.50 & $31 \pm 14$ & $0.1 \pm 0.3$ & $1.0^{+2.3}_{-0.5}$ & 12.8$^{+0.9}_{-0.8}$ & 14.2$^{+3.2}_{-4.4}$ & $896^{+218}_{-157}$ & 0.62 (41) \\
12 & 0.55 & $12 \pm 6$ & $0.5^{+0.2}_{-0.3}$ & $1.8^{+1.6}_{-0.9}$ & 15.2$^{+2.3}_{-2.0}$ & 21.3$^{+8.8}_{-10.2}$ & $675^{+138}_{-124}$ & 1.04 (41) \\
13 & 0.60 & $14^{+4}_{-3}$ & $1.3 \pm 0.1$ & $10.9^{+3.5}_{-3.6}$ & 20.5$^{+2.7}_{-8.2}$ & 5.0$^{+63.0}_{-4.9}$ & $196^{+82}_{-98}$ & 0.91 (41) \\
14 & 0.65 & $22^{+23}_{-13}$ & $1.1^{+0.7}_{-0.5}$ & $4.5^{+26.1}_{-3.3}$ & 16.2 & 22.0 & $<638$ & 0.58 (43) \\
15 & 0.70 & $30^{+6}_{-5}$ & $1.3 \pm 0.2$ & $15.8^{+4.0}_{-5.9}$ & 16.2 & 22.0 & $<202$ & 1.10 (43) \\
16 & 0.76 & $16^{+14}_{-13}$ & $1.5 \pm 0.3$ & $6.9^{+2.3}_{-0.5}$ & $17.6 \pm 1.4$ & 17.0$^{+13.2}_{-14.4}$ & $<380$ & 0.61 (41) \\
17 & 0.81 & $25^{+8}_{-6}$ & $1.3 \pm 0.2 $ & $9.1^{+9.0}_{-5.5}$ & 16.2 & 22.0 & $<181$ & 0.82 (43) \\
18 & 0.86 & $46^{+20}_{-12}$ & $1.7^{+0.5}_{-0.2}$ & $20^{+70}_{-13}$ & 16.2 & 22.0 & $<232$ & 0.85 (43) \\
19 & 0.91 & $126^{+84}_{-60}$ & $1.3^{+0.7}_{-0.6}$ & $3.8^{+175.4}_{-2.2}$ & 16.2 & 22.0 & $<6490$ & 0.94 (43) \\
\enddata
\tablecomments{XSPEC model: {\tt phabs*(raymond + power) + phabs*highecut*(power + gauss)}. All errors are quoted at the 90\% confidence level for a single parameter. The Galactic ridge emission parameters were fixed at the best-fit values of the simultaneous fit to spectra numbers 1, 2, \& 20, which occurred during eclipse (see Table 4). Model normalizations follow the definitions used in XSPEC (Arnaud 1996).}
\tablenotetext{a}{The orbital phase at the middle of the observation, inferred from the fit presented in \S~3.1.} 
\tablenotetext{b}{The high energy cut-off and folding energies were allowed to vary only if an {\em F}-test showed that there is less than a 5\% probability that the improvement to the fit due to the cut-off occurred by chance. In all other cases, the cut-off and folding energies were fixed to the best-fit value found by fitting all spectra simultaneously with a single cut-off and folding energy (see text).}
\tablenotetext{c}{Equivalent width of narrow Fe line at 6.4 keV.}
\end{deluxetable*}

\section{Discussion of the Spectrum}
\begin{figure}
\centering
\includegraphics[width=3.5in]{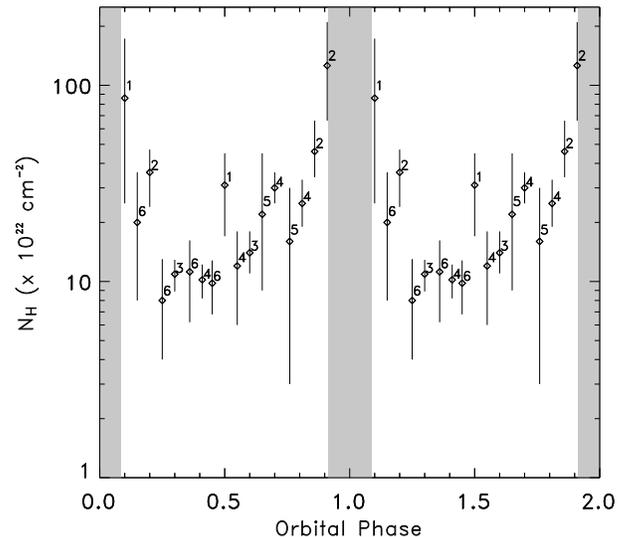}
\caption{Best-fit hydrogen column density (Table 5) as a function of orbital phase. The orbit from which the $N_{\rm H}$ measurement was taken is labeled to the right of each point. The shaded bar indicates times when the pulsar is eclipsed by the primary.\label{nh}}
\end{figure}

The orbital modulation of the EXO 1722-363 spectrum, and especially the hydrogen column density, is intriguing. Figure \ref{nh} shows $N_{\rm H}$ plotted against orbital phase for two cycles; a vertical shaded bar shows when the pulsar is eclipsed by the primary. The two observations with the highest column densities occur just prior to ingress and just after egress, at a point where the pulsar is viewed through the base of the stellar wind at the very edge of the primary. Increased absorption for these observations is also supported by the abrupt increase in the hardness ratio near eclipse (see Fig. \ref{orbmod}). Between phases 0.25--0.45, $N_{\rm H}$ is consistent at about $10^{23}$ cm$^{-2}$, but apparently increases for the second half of the orbit. The error bars are somewhat large, but at phase 0.70 the increase in $N_{\rm H}$ is significant. It is difficult, however, to state whether we are observing variability in the column density due to differences within the orbit or differences between orbits. The specific orbit where each measurement was made is listed next to its corresponding data point in the figure. 

Phase-dependent modulation of the hydrogen column density has been investigated by several authors by studying the influence of a compact object on the stellar wind of the primary through observations and hydrodynamic simulations (e.g., Haberl et al. 1989; Blondin et al. 1990; Blondin et al. 1991; Blondin 1994). Stellar wind flowing past a compact object is affected by a number of competing forces (gravitational, rotational, and radiation pressure), and photoionization of the stellar wind can create an accretion wake (Fransson \& Fabian 1980). In addition, if the orbital separation is small enough, tidal distortion of the surface of the primary can lead to enhanced mass loss through the inner Lagrangian point $L_{1}$ along the line of centers even if the primary does not fill its Roche lobe (Friend \& Castor 1982). The presence of a tidal stream depends sensitively on the separation of the primary and the compact object, and only becomes apparent when the radius of the primary is greater than 95\% of the radius of its Roche lobe (Blondin et al. 1991). Even if a tidal tream is not present, the simulations by Blondin et al. (1990) are described by the formation of nonsteady accretion wakes that produce dense filaments in the downstream wake. If the filaments are along the line-of-sight to the pulsar, which is likely to occur at orbital phases 0.5--1.0, the observed hydrogen column density will increase by a factor of a few to 10, qualitatively consistent with the phase-dependent photoelectric absorption seen in \igr.

Models including a significant tidal stream show weak or altogether absent changes in the sign of the accreted angular momentum (Blondin et al. 1991). While the connection of pulses from 2003 August to 2006 March using a single pulse period derivative is indicative of a stable accretion process over the 2.4 yr span (spinning up the neutron star), the orbital solution unambiguously shows that at least one torque reversal occurred between 1998 November and 2003 August: The pulse period increased from 413.85 s to 414.65 s between late 1998 and 2003 August, but has been steadily decreasing through early 2006. Note that the low-energy light curve of EXO 1722-363 showed no dips, and so there is no evidence for a persistent accretion disk. The spin-down therefore occurred sometime between 1998 November and 2003 August. Such spin-down can occur via the ``propeller effect" (Illarionov \& Sunyaev 1975), or if there are significant azimuthal gradients in the wind density or velocity near the compact object (Wang 1981), however, as mentioned above, a torque reversal is inconsistent with the simulations predicting increases in absorption at later phases due to a dominant tidal stream. 

Observations of phase variability of the column density in Vela X-1 and 4U 1700-37 qualitatively agree with that seen in EXO 1722-363, with significantly higher absorption after $\Phi_{\rm orb}=0.5$ (e.g., Haberl 1989; Haberl et al. 1989; Charles et al. 1978).  A base level column density is seen to be reproducible over many orbits, along with a highly variable component causing $N_{\rm H}$ to increase by an order of magnitude over the base level, usually appearing at later phases. This variable component is suggested by Blondin et al. (1991) to result from separate clumps of material formed by dynamical instabilities in the accretion bow shock. Interestingly, one of the narrow peaks in Vela X-1 consistently appears at phase 0.5, and we see similar behavior, at least for one particular orbit, in EXO 1722-363. 

The picture that emerged above is one of the pulsar moving through the stellar wind of its B-type companion, affecting the distribution of matter through a number of competing processes, and possibly leading to a wake of dense material trailing the pulsar in its orbit. This material will also reprocess the radiation incident upon it, leading to fluorescent iron line emission. The interpretation of the variability of the Fe K$\alpha$ line is complicated because the equivalent width of the line depends on many factors, such as the spectrum of the continuum emission, and the distribution and density of the fluorescing material. The results of Table 5 show that the photon index varies over a wide range throughout the orbit (the photon index $\Gamma$ is consistent with values between -0.5 and 2.7 at the 90\% confidence level), as does the hydrogen column density. The distribution of the surrounding matter is also expected to vary over the orbit, given the nature of stellar winds from massive stars and the effect of the pulsar on these winds.
\begin{figure*}
\centering
\includegraphics[width=5.5in]{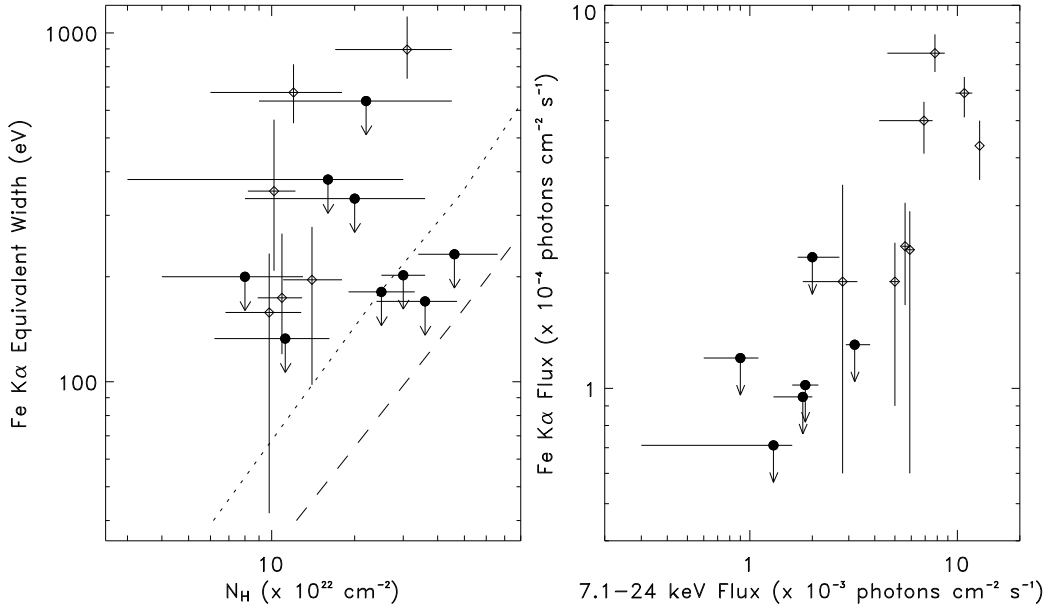}
\caption{Column density versus the equivalent width of the iron line ({\em left panel}), and the 7.1--24 keV continuum intensity versus the iron line flux ({\em right panel}). Data points representing upper limits are indicated with closed circles, and the non-upper limit data points are indicated with open diamonds. The error bars in the left panel represent the 90\% confidence level, while the error bars in the right panel represent the 68\% confidence level. The data points from spectra 3 and 19, which have extremely large upper limits to the equivalent width, are not shown in either panel. The {\em dotted} line in the {\em left} panel shows the predicted relation between $N_{\rm H}$ and the Fe K$\alpha$ equivalent width for power-law radiation with $\Gamma=1$ emerging from a uniform sphere surrounding the source, and is the result of simulations by Leahy \& Creighton (1993). The {\em dashed} line is the same but for $\Gamma=2$. \label{iron}}
\end{figure*}

Using Monte Carlo simulations, Leahy \& Creighton (1993) found that an input spectrum with $\Gamma = 1$ emerging from a uniform and spherically symmetric distribution of matter will lead to an iron line equivalent width of about 100 eV for a column density of $10^{23}$ cm$^{-2}$. These authors also found that equivalent width increases approximately linearly below a column density of $10^{25}$ cm$^{-2}$. An input spectrum with $\Gamma = 2$, on the other hand, results in equivalent widths about one-third as large, simply because a spectrum described by a steeper power law has relatively fewer photons above the iron K-edge. Figure \ref{iron} shows the column density versus the equivalent width of the iron line ({\em left} panel), and the 7.1--24 keV continuum intensity versus the iron line flux ({\em right} panel). Unfortunately, upper limits on the iron line intensity were obtained for more than half of the spectra (closed circles in the figure). Also, the data points from spectra 3 and 19 were not included in the figure because the flux for these observations is only $\sim$10\% above the GRE emission, and poor statistics lead to very weakly constrained equivalent widths. If we only consider the data points that are not consistent with zero (open diamonds in the figure), one can see a rough correlation between the Fe K$\alpha$ equivalent width and the $N_{\rm H}$ column density, but the magnitude of the equivalent width is systematically larger than that predicted by the Leahy \& Creighton (1993) simulations. Perhaps this reflects an iron abundance greater than the solar value. The dependence of the equivalent width on the iron abundance is linear when the optical depth is much less than one, but becomes less than linear when the effects of opacity become important (Matt 1997). Nevertheless, by including data points that are upper limits, this rough trend no longer exists. One factor that may contribute to the lack of a clear correlation is the significant variability of the photon index. The three largest values for the equivalent width (disregarding upper limits) are from the three spectra with the flattest power laws (best-fit photon indexes of $\Gamma =0.1$, 0.5, \& 0.9, respectively), as we would expect. The right panel of Fig. \ref{iron} shows that the correlation between the iron line flux and the continuum intensity above the iron K-edge is better, though still not excellent. If the correlation were linear and the spectrum constant (minus changes in absorption), it would indicate that the origin of the iron line is from the same material as is responsible for the absorption of the continuum. Spectral variability makes such a comparison impossible, but we speculate that the pulsar may be illuminating matter that is not along its line of sight, and that may have a smaller . That the equivalent width increases near $\Phi_{\rm orb}=0.5$ possibly indicates that a component of the fluorescing material is the surface of the primary.

\section{Summary and Conclusions}
We have found the orbital solution to the eclipsing HMXB \igr~by applying pulse timing analysis to twenty recent {\em RXTE} observations (nearly evenly spaced in orbital phase) and two sets of observations occurring 2.4 years and 7.3 years earlier.  With the orbital solution, we constrained the nature of the mass-donating companion star, and we investigated the evolution of the spectrum through the pulsar's orbit. Our primary results are summarized as follows:

\begin{enumerate}

\item 
The circular orbital solution is characterized by $a_{x} \sin{i} = 101 \pm 3$ lt-s, $P_{\rm orb} = 9.7403 \pm 0.0004$ days, and an epoch of mid-eclipse at HJD 53761.68 $\pm$ 0.04. The eccentricity is less than 0.19 at the 90\% confidence level. Our orbital solution also implies that a reversal of the torque on the neutron star occurred between 1998 November and 2003 August. 

\item
Using the orbital solution and the eclipse half-angle measurement of $\theta_{e} = 31\fdg8 \pm 1\fdg8$ from Corbet et al. (2005), we constrained the companion star's mass and radius as a function of inclination. Assuming a neutron star mass of 1.4 $M_{\sun}$ and that the primary underfills its Roche lobe and, the system must have an inclination greater than 61\degr~with 99\% confidence. With the lower limit on the inclination, the primary's mass is between about 11 $M_{\sun}$ and 22 $M_{\sun}$, and its radius is between 21 $R_{\sun}$ amd 37 $R_{\sun}$. The range of acceptable masses and radii are consistent with a supergiant star with a spectral type between B0 and B5. 

\item
There is only one 2MASS IR counterpart in the 4\arcsec~{\em XMM-Newton} error circle. Assuming this source is the primary, we found that the photometric measurements from 2MASS and DENIS are consistent with the inferred spectral type if $A_{V}=20.1 \pm 0.6$ and the distance to the system is between 5.3 kpc and 8.7 kpc. In the 2--24 keV band, the unabsorbed X-ray luminosity is $L_{x} \sim$ 0.2--4.4 $\times 10^{36}$ ergs s$^{-1}$.

\item 
The spectrum of \igr~shows variable flux, absorbing column density, power-law photon index, and iron line equivalent width through the course of the orbit. Between orbital phases 0.25--0.45, $N_{\rm H}$ is consistent at about $10^{23}$ cm$^{-2}$, but apparently increases for the second half of the orbit.

\end{enumerate}

\acknowledgements
We gratefully acknowledge support from NASA grants NNG06GA85G and NAG5-30720. We thank Craig Markwardt and Keith Jahoda for assisting us in accounting for uncertainties in the PCA background models. We also thank R\"{u}diger Staubert for guidance in creating phase-connected datasets. This research has made use of data obtained through the High Energy Astrophysics Science Archive Research Center Online Service, provided by the NASA/Goddard Space Flight Center.


\end{document}